\documentclass[sigconf]{acmart}
 \usepackage{amsmath,amsfonts}
\let\chapter\section           
\usepackage{textcomp}
\usepackage{xcolor}
\usepackage{wrapfig}
\usepackage{color,soul}
\usepackage{multirow}
\usepackage{url}
\usepackage{tikz}
\usepackage{comment}
\usepackage{subcaption}
\usepackage{graphicx}
\usepackage[absolute]{textpos}
\usepackage[flushleft]{threeparttable}
\usepackage[linesnumbered,ruled,vlined]{algorithm2e}
\usepackage{algpseudocode}

\newlength\myindent
\AtBeginDocument{%
  \providecommand\BibTeX{{%
    \normalfont B\kern-0.5em{\scshape i\kern-0.25em b}\kern-0.8em\TeX}}}

\begin{document}

\title{From Zero to The Hero: A Collaborative Market Aware Recommendation System for Crowd Workers}

% \author{Hamid Shamszare}
% \email{hshamsza@stevens.edu}
% \affiliation{%
%   \institution{Stevens Institute of Technology}
% %   \streetaddress{P.O. Box 1212}
%   \city{Hoboken}
%   \state{New Jersey}
% %   \postcode{43017-6221}
% }

\author{Hamid Shamszare}
\authornote{Both authors contributed equally to this research.}
\affiliation{%
  \institution{Stevens Institute of Technology}
  \city{Hoboken, NJ}
  \country{USA}}
\email{hshamsza@stevens.edu}

\author{Razieh Saremi}
\authornotemark[1]
\affiliation{%
  \institution{Stevens Institute of Technology}
  \city{Hoboken, NJ}
  \country{USA}}
\email{rsaremi@stevens.edu}

\author{Sanam Jena}
\affiliation{%
  \institution{Stevens Institute of Technology}
  \city{Hoboken, NJ}
  \country{USA}}
\email{sjena@stevens.edu}

% \author{Ye Yang}
% \affiliation{%
%   \institution{Stevens Institute of Technology}
% %   \streetaddress{1 Th{\o}rv{\"a}ld Circle}
%   \city{Hoboken}
%   \state{New Jersey}
% \email{yyang4@stevens.edu}}

\renewcommand{\shortauthors}{Shamszare, et al.}

\begin{abstract}
The success of software crowdsourcing depends on active and trustworthy pool of worker supply. The uncertainty of crowd workers' behaviors makes  it challenging to predict workers' success and plan accordingly. In a competitive crowdsourcing marketplace, competition for success over shared tasks adds another layer of uncertainty in crowd workers' decision making process. Preliminary analysis on software worker behaviors
reveals an alarming task dropping rate of 82.9\%.  
These factors lead to the need for automated recommendation system for CSD workers to improve the visibility and predictability of their success in the competition. 
To that end, this paper proposes a collaborative recommendation system for crowd workers.
The proposed recommendation system  method uses five input metrics based on workers collaboration history in the pool, workers preferences in taking tasks in terms of monetary prize and duration,  workers' specialty, and workers' proficiency. The proposed method then recommends the most suitable tasks for a worker to compete on based on workers' probability of success in the task. Experimental results on 260 active crowd workers demonstrate that just following the top three success probability of task recommendations, workers can achieve success up to 86\%.

\end{abstract}




\maketitle

\section{Introduction}
Crowdsourced software development (CSD) integrates online and unknown workers’ elements into the design. The success of such platforms relies heavily on a large crowd of trustworthy software workers who are registering and submitting for crowdsourced tasks in exchange of financial gains \cite{yang2016should}. 
Crowd workers make their decision on participating in a task based on intrinsic values which result from workers' motivation and preferences \cite{si2014encoding} \cite{leimeister2009leveraging}, and extrinsic values on the societal level such as educational background and household income\cite{kaufmann2011more}.

In general, crowd workers choose to perform tasks based on some personal utility algorithm such as monetary prize and tasks' complexity and duration \cite{yang2015award}\cite{saremi2015dynamic}\cite{mejorado2020study}, their skills and some unknown factors \cite{faradani2011s}. However, they rather to work on the tasks with similar context in terms of task type, required technology and platform, and their previous experience \cite{difallah2016scheduling}\cite{crump2013evaluating} \cite{yin2014monetary}.

It seems that crowd workers analyze their probability of success based on the task’s competition level and the number of more highly ranked opponents \cite{faradani2011s}, and use their history of victories as a policy of assuring to win the registered tasks \cite{faradani2011s} \cite{archak2010money}. But, crowd workers often overestimate their productivity \cite{jorgensen2005over}, and they register for more tasks than they can complete. It is reported that 82.9\% of active worker in topcoder\footnote{ \url{https://www.topcoder.com/}} drop their registered tasks, and 55.8\% of the submissions are not valid \cite{yang2016should}.  Therefore, understanding opponent performance and trustworthiness is essential for crowd workers' decision-making processes.

The objective of this study is to propose a collaborative recommendation system to support crowd workers dynamic decision making to explore  options and improve  their probability of success in CSD. To this end, we first present a motivational example to highlight workers relation in CSD platform. Then we propose a collaborative recommendation system . This system seeks to increase workers probability of success in competing on different tasks in the platform. The system takes as input a list of open tasks in the platform, workers' collaboration and attributes and previous performances. It then recommends the list of suggested tasks to workers based on workers' specialty, and proficiency and probability of success. 

The input of the proposed system  is conducted on more than one year’s real-world data from topcoder, the leading software crowdsourcing platform with an online community of over 1.5M workers and 55k mini-tasks . We applied the proposed recommendation system to 260 active crowd workers in our database during the weeks of Jan 14\textsuperscript{th} 2015 to Jan 30\textsuperscript{th} 2015. For workers just following the top three task success probability, workers can achieve success up to 86\%.

The remainder of this paper is structured as follows. Section 2 introduces a motivational example that inspires this study. Section 3 presents background and related work. Section 4 outlines our research design. Section 5 discuses the results of this research. Section 6 presents the conclusion and outlines a number of directions for future work.

\section{Motivating Example} \label{example}

\begin{figure}[ht!]
\centering
\includegraphics[width=0.9\columnwidth]{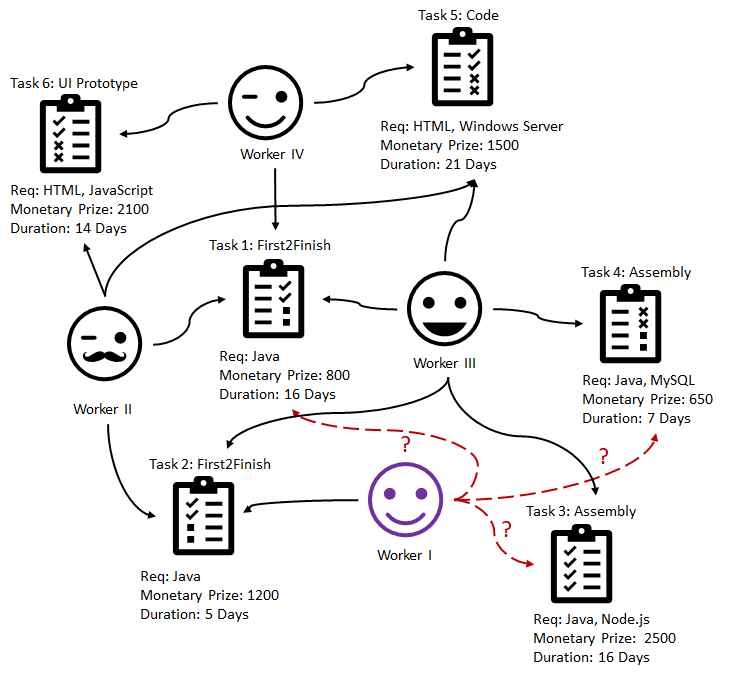}
\caption{Overview of Motivating Example}
\label{motivating}
\end{figure}

Figure \ref{motivating} depicts a motivating example with task selection, task requirement, task type and competition information among four workers and six tasks. The information on top of each task are task ID and task type and the information under each task are technical requirements to perform the task, Monetary Prize associated with tasks and task duration. All of the six tasks opened for competition in the same week when the four workers (i.e worker I, II, II, and IV) were available to take tasks at the same time. Table \ref{workers} summarises the four workers profile and previous performance.
As it is illustrated in figure \ref{motivating}, Worker II registered for four tasks to compete on (i.e. tasks 1,2,5 and 6). These tasks are chosen from different tasks types of First2Finish, Code, and UI Prototype. Also they required combination of different technologies such as Java, HTML, and Java Script. 
Worker III registered for five tasks  of 1, 2, 3, 4, and 5. Two of First2Finish, two of Assembly and one of Code type. While four of these tasks can be performed with knowledge of Java, the other one (i.e. task 5) requires HTML or Windows Server to be done. 
At the same time Worker IV  registered for tasks 1, 5 and 6. While tasks 5 and 6 can be performed only by knowing HTML, task 1 requires some  knowledge of Java. Furthermore, the three tasks belong to three different task types: First2Finish, Code, and UI Prototype. 
Interestingly, Worker I only registered for task 2, which requires Java and is under First2Finish type. 

\textit{Is there any other task that Worker I can confidently register for?}

The relation among the four workers and six tasks creates a network. In this network, Worker II and Worker III are common neighbors for Worker I. This means tasks that these workers registered for can be a potential task for Worker I to take. Given the similarity in Worker I's \textit{proficiency} in Java, according to his previous registration based on table\ref{workers}, most probably tasks 1,3, and 4 are good fit for him to work on. Given his \textit{specialty} in First2Finish, task 1 could be number one match for Worker I to compete on.  

To register for a task, workers make the decision based on there previous performances, types of tasks they are interested, skill match to those required by the task, affordability in terms of time and effort required to complete a task, associate monetary prize, and reputation of other competitors on the task. 

In this example, task 1 meets worker I preferences in terms of monetary prize, duration, specialty and proficiency (table \ref{workers}). However, taking task 1 means that worker I will compete against worker II, worker III, and worker IV. knowing competitors preferences and performance history would enable worker I to predict his/her success level before make a decision on taking task 1.

This observation motivated us to propose a worker recommendation method based on a combination of collaborative filtering systems and workers individual preference and performance. This method can increase the probability of workers' success in performing tasks in the platform.

\section{Background and Related Work}

\subsection{Workers Performance in Crowdsourcing}

Software workers’ arrival to the platform and the pattern of taking tasks to compete on, are factors that shape the worker dynamic in a crowdsourcing platform, however, the reliability in returning the qualified tasks creates the dynamic of the platform. Generally, not only would the monetary prize associated with the task influence the workers’ interests in competitions\cite{yin2014monetary}, the number of registrants for the task, the number of submissions by individual workers, and certainly the workers’ historical score rate would directly affect their final performance \cite{lakhani2010topcoder}\cite{saremi2017leveraging}. For newcomers or beginners, there is a time window required to improve and to develop into an active worker \cite{faradani2011s}. Therefore, it is typical that the workers need to communicate with the task owner in order to better understand the problems to be solved \cite{kittur2013future}. Existing studies show that over time, registrants gain more experience, exhibit better performance, and consequently gain higher scores \cite{faradani2011s} \cite{archak2010learning} \cite{kittur2013future}. Still, there are workers who manage not only to win but also to raise their submission-to-win ratio \cite{difallah2016scheduling}. This motivate workers to develop behavioral strategies in topcoder \cite{archak2010money} \cite{archak2010learning}. Moreover, the ranking mechanism used by topcoder contributes to the efficiency of online competition and provides more freedom of choice for higher rate workers in terms of controlling competition level \cite{archak2010money}.

\subsection{Workers Competition in Crowdsourcing}

To encourage contribution and engagement, crowdsourcing adopt the use of extrinsic rewards such as ranking as game elements for workers to compete in a non-gaming context \cite{cheng2020building}. Extrinsic rewards can increase the overall workers' engagement and commitment\cite{bista2012using}\cite{cavusoglu2015can}, motivation \cite{CHANDLER2013123}\cite{soliman2015understanding}\cite{stewart2009designing} and collaboration \cite{gray2016crowd}, since they address a type of social need for some community members\cite{richter2015studying}.

In a CSD platform, a competitive environment not only influences the decisions of workers regarding which tasks to register and submit but also how they react to their peers. Such environment creates opportunities for workers to apply different strategies and assure their success and increasing their rank in the system.
One primary example is rank-boosting \cite{ye2015crowd}\cite{ipeirotis2010top} in Amazon Mechanical Turk, where workers mostly register for easy tasks or fake tasks that they themselves are uploading in order to increase their rating. Another example is detecting cheap talk phenomena \cite{farrell1996cheap}\cite{archak2010money} in topcoder. In CSD, higher rated workers have more freedom of choice in comparison with lower rated workers and can successfully affect the registration of lower rated workers. To assure a softer and easier competition level, higher rated workers are more likely to launch challenges against lower ones\cite{hu2014game}.

\subsection{Workers Decision Making in Crowdsourcing}

Online decision algorithms have a rich literature in operations
research, economics, machine learning, and artificial intelligence,
etc. Much of existing work on crowdsourcing decision making focuses on assigning reliable workers to existed task, such as learning worker quality and optimizing task assignment decisions \cite{whitehill2009whose}, aggregating individual answers to improve quality
\cite{mao2013better}, and worker incentives \cite{kaufmann2011more}, developer recommendations \cite{mao2015developer}\cite{difallah2013pick}, and
understanding worker behaviors \cite{yang2015award}\cite{archak2010money}\cite{zhang2015analyzing}.
In software crowdsourcing, only a few studies have focused on
decision support for software crowdsourcing market. Among them, Mao et al. \cite{mao2015developer} presented a content-based developer recommendation
framework for CSD context, to recommend reliable workers based
on static features extracted from participation history and winning
history. Difallah et al.\cite{difallah2013pick} propose a recommendation framework to pick a suitable crowd for a task. Yang et al. \cite{yang2016should} introduces 'DCW-DS', an analytic-based decision support methodology to guide workers
in acceptance of offered development tasks. Hettiachchi et al. \cite{hettiachchi2020crowdcog} present 'CrowdCog',an online dynamic system that provides both task assignment and task recommendations, based on online cognitive tests to estimate worker performance across a variety of tasks. And Kumar et al.\cite{abhinav2020tasrec} proposed 'TasRec', a worker’s fitment framework based on worker’s preference, past tasks (s)he has performed, and tasks done by similar workers.

\begin{table*}[!ht]
\caption{Summary of Metrics Definition} % title of Table
\centering % used for centering table
\begin{tabular}{p{2cm} p{4cm} p{10cm}}
\hline
Type & Metrics & Definition \\ %[0.5ex] % inserts table
\hline%\hline % inserts single horizontal line
\multirow[origin=c]{8}{*}{\parbox{1.5cm}{Workers   Attributes}} 
& \# Registration (R)  & Number of registrants that are willing to compete on total number of tasks in specific period of time.  \\ %Range: (0, $\infty$).
& \# Submissions (S) & Number of submissions that a task receives by its submission deadline in specific period of time. \\ % Range: (0, \# registrants).
& \# Valid Submissions (VS) & Number of submissions that a task receives by its submission deadline and passed the peer review and labeled as either completed or active.\\ 
% & Win (W) & Number of submissions that pass the peer review and labeled as active submission.\\
\hline
\multirow{4}{*}{\parbox{1.5cm}{Workers-Tasks Attributes}}
& Worker registration date (WR) & Date and time that a worker registered for a task. (mm/dd/yyyy) \\
& Worker submission date (WS) & Date and time a worker submitted for a task. (mm/dd/yyyy) \\
& Duration (D) & total available days from worker registration date ($ {TR} $) to submissions date ($ {TS} $) \\
& Task Status & Completed or failed task. \\
& Task Type (Type) & Type of task competition.\\
& Task Competition Level (TC) & Total number of workers to register and are willing to compete on a task. \\
& Monetary Prize (MP) & Monetary prize (USD) offered for completing the task and is found in task description. Range: (0, $\infty$) \\
% & \# Starved Tasks (ST) & Number of Tasks that receives zero submission by its submission deadline and failed. \\
& Technologies (Tech)  & Number of technologies used in task.  \\ %Range: (0, $\infty$).
& Task Duration (D) & total available days from task registration start date ($ {TR} $) to submissions end date ($ {TS} $) \\
% & Platform (PLT) & Number of platforms used in task. \\ % Range: (0, \# registrants).
% & Worker expertise (WTech) & Number of technologies used in tasks that worker compete on.\\ % Range: (0, \# registrants).
% & Worker expertise (WTech) & Number of technologies used in tasks that worker compete on.\\ % Range: (0, \# registrants).

\hline
\label{metrics}
\end{tabular}
\end{table*}

\section{Research Design} \label{model}

To develop a recommendation model from workers perspective, we use collaborative filtering method to increase workers proficiency and specialty in the suggested tasks. Then, we provide the probability of success per recommended task and report top three tasks with highest success probability per worker. This helps workers to have a higher confidence in making a decision to take the tasks. 
% In this paper, we only report top three tasks with highest success probability per worker. 
This architecture can be applied to any crowdsourcing platform; however, we focused on Topcoder as the target platform. 

\begin{figure}[ht!]
\centering
\includegraphics[width=0.9\columnwidth]{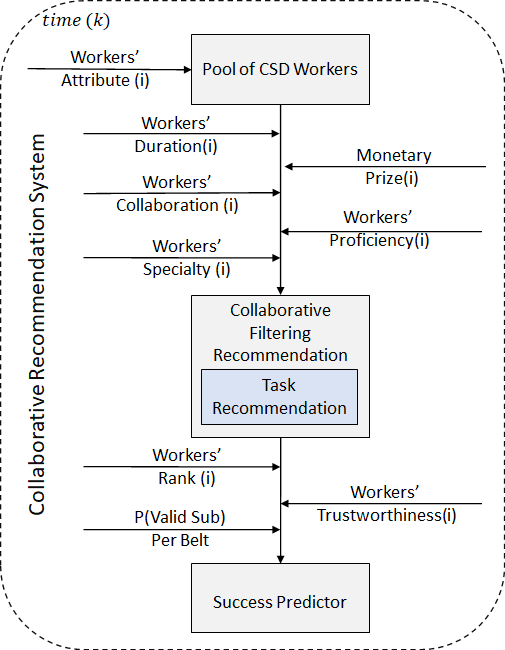}
\caption{Overview of Collaborative Recommendation System Architecture}
\label{diagram}
\end{figure}

In this method, most suitable open tasks during two weeks from any point of time are suggested based on the degree of compatibility to workers' preferences in terms of monetary prize, duration, and workers' skill set in terms of  proficiency, and specialty level; then, a worker could choose to compete in the proper tasks based on the probability of success in making a valid submission for the worker and the other registrants of the tasks.  Figure \ref{diagram} presents the overview of the collaborative recommendation system architecture. Workers' attributes are uploaded in the \textit{pool of workers}. From this pool, \textit{workers' rank} based on Topcoder definition\cite{saremi2017leveraging}, workers' preferences in terms of \textit{monetary prize}, \textit{duration}, \textit{workers' collaboration}, \textit{workers' specialty}, \textit{workers' proficiency}, and \textit{workers' trustworthiness} are calculated. The calculated metric is used as an input to \textit{collaborative filtering system} which provides a list of \textit{recommended tasks} to the workers as an output. Then, \textit{worker success predictor} analyzes workers' probability of success per recommended task and ordered the top three recommendations.

\subsection{Crowd Workers} 
According to Howe \cite{howe2008crowdsourcing}, crowd workers are a large and undefined group of skilled workers who have access to a task via the internet. In this research, we followed the definition of CSD workers in \cite{saremi2021crowdsim}. 
% CSD workers (${W_{i}}$)  are a tuple of different characteristics of workers’ identification (${AID_{i}}$), reliability factor (${Re_{i}}$), rating (${Ra_{i}}$), skillset (${SK_{i}}$), score (${So_{i}}$), number of competitions (${R_{i}}$), number of submissions (${S_{i}}$), number of valid submissions (${VS_{i}}$), number of wins (${Wi_{i}}$), location (${L_{r}}$), and his/her membership tenure (${MA_{r}}$) at a given time t (${t_{q}}$). A crowd worker is defined with the following formula: 
% \begin{equation}
% {W_{i}} = {({AID_{i}}, {Re_{i}}, {Ra_{i}}, {SK_{i}}, {So_{i}}, {R_{i}}, {S_{i}}, {VS_{i}}, {Wi_{i}}, {L_{i}}, {MA_{i}})}
% \end{equation}
% \[
% where
%   \text{  i = 1, 2, 3, …, r} 
% \]

\subsubsection{Workers' Attributes}
Worker attributes are the subset of workers characteristics that dynamically influences successful task delivery per community. The list of Worker attributes applied in this study is summarized in Table \ref{metrics}.

\subsubsection{Workers' Monetary Prizes}
The minimum monetary prize, (${bMP_{i}}$), associated to the list of tasks performed by Worker(i), ${W_{i}}$, is used as the base monetary prize for the recommended tasks to the worker.

\begin{equation}
{bMP_{i}} = {\min_{i=0}^{n}{(MP_{i})}}
\end{equation}

\subsubsection{Workers' Duration }
Total available days from worker registration date (${WR_{i}}$)to submissions date (${WS_{i}}$) is the duration (${D_{i}}$). In this research, the minimum value in list of duration periods worker $i$ has in his/her profile is used as the base duration,(${bD_{i}}$), for the recommended task. 

\begin{equation}
{D_{i}} = {({WR_{i}} - {WS_{i}})}
\end{equation}

\begin{equation}
{bD_{i}} = {\min_{i=0}^{n}{(D_{i})}}
\end{equation}

\begin{table}[!ht]
\caption{Summary of Different Workers' Rank in Topcoder} % title of Table
\centering % used for centering table
\begin{tabular}{p{2.5cm} p{0.7cm} p{2cm} p{1cm} p{1cm}}
\hline
\\
Workers' Belt & & Rating Range(X) &  Workers\% & p(VS)\\ %[0.5ex] % inserts table

\hline%\hline % inserts single horizontal line

\multirow{3}{*}{Lower Experienced} 
\\
& Gray &  $ {X < 900} $ & 90.02\% & 0.25 \\
\\
\hline

\multirow{5}{*}{Average  Experienced} 
\\
& Green & ${900 < X < 1200}$ & 2.88\% & 0.45 \\
\\
& Blue & ${1200 < X < 1500}$ & 5.39\% & 0.39 \\
\\
\hline

\multirow{5}{*}{Higher Experienced} 
\\
& Yellow & ${1500 < X < 2200}$ & 1.54\% & 0.6\\
\\
& Red & ${X > 2200 }$ & 0.16\% & 0.6 \\
\\
\hline
\label{belt}
\end{tabular}
\end{table}
%%%%%%%%%%%%%%%%%%%%%%%%%%%%%%%%%%%%

\subsubsection{Workers' Rank} 
Topcoder adopts a  numeric worker rating system based on Elo rating algorithm\footnotemark\footnotetext{\url{https://www.topcoder.com/member-onboarding/understanding-your-topcoder-rating}}, as a result topcoder workers are divided into 5 groups. The 5 worker groups are defined as 5 belts of Red, Yellow, Blue, Green, and Gray, which corresponds to the highest skillful workers
to the lowest ones\cite{saremi2017leveraging}. 
Table \ref{belt} summarizes the distribution of workers belonging to different rating belts, (${WB}$), in the Topcoder dataset used in this study. It is shown that among the total of 5062 active workers, more than 90\% of the workers are in Gray belt, which is the non-experienced group. The other 10\% of workers are more experienced solid workers.

% \subsubsection{Workers' Collaboration}

% The history of two workers competing on the same task creates workers' collaboration (${WC_{i,j}})$.
% We create binary variables for each pair of workers $i$ and $j$, ${W_{i}}-{W_{j}}$, based on their registration history for different task, where:
% \begin{equation}
% {WC_{i,j}} =
% 	\begin{cases}
%         1 &\parbox[t]{.35\textwidth}{${W_{i}}$ and ${W_{j}}$ registered for the same task} \\
%         0 & \text{otherwise}
%     \end{cases}
% \end{equation}

% \hl{I need help with..}
% \subsubsection{Workers' Proficiency}
% To  understand workers' proficiency in performing a tasks we need to understand workers' skill match to those required by a task. Therefor,worker proficiency level ${PL_{i}}$  defines as the probability of worker$i$ registered for a task with similar technology requirements ($WTech$) in workers' history of performance.

% \begin{equation}
% {PL_{i}} = \frac{\sum_{i=0}^{n}{WTech_{i}}} 
% {\sum_{i=0}^{n}{R_{i}}}
% \end{equation}

% \subsubsection{Workers' Specialty}
% Workers' specialty level ${SL_{i}}$ in performing a task is defined as the probability of worker$i$ registered for a task from the similar task type ($WType$) in workers' history of performance.

% \begin{equation}
% {SL_{i}} = \frac{\sum_{i=0}^{n}{WType_{i}}} 
% {\sum_{i=0}^{n}{R_{i}}}
% \end{equation}

\subsubsection{Workers' Proficiency}
To  understand workers' proficiency in performing a task, we need to realize if the worker's skills match to those required by the task or not.
% Therefor,worker proficiency level ${PL_{i}}$  defines as the probability of worker$i$ registered for a task with similar technology requirements ($WTech$) in workers' history of performance.
% Worker proficiency level ${PL_{i}}$ is an indicative of the preferences of worker $i$ for the technology ($WTech$) required by the recommended task. I is calculated as the ratio of the frequency of worker $i$’s participation in the technology $r$ (${WTech_{r}}$)  to all teh technologies worker $i$ has used.
% and …Is  the frequency of the worker i’s participation in all of the technologies...
% Is  the frequency of the worker i’s participation in all of the technologies....... the formula is available in the word version}

Worker proficiency level ${PL_{i}}$ is indicative of the preferences of worker $i$ for the set of technologies required by the recommended task. For a specific technology  $t$, the proficiency is calculated as the ratio of the frequency of worker $i$’s participation in the technology $t$, (${WTech_{i,t}}$) , to the frequency of all the technologies worker $i$ has used (${WTech_{i,j}}$).

\begin{equation}
{PL_{i,t}} = \frac{{WTech_{i,t}}} 
{\sum_{j=1}^{n}{WTech_{i,j}}}
\end{equation}

When Worker i's proficiency is calculated based on $N$ different technologies required by task $r$, average value of ${PL_{i,r}}$ is utilized as workers' proficiency.

\begin{equation}
{APL_{i,r}} = \frac{\sum_{t=1}^{N}{PL_{i,t}}} 
{{N}}
\end{equation}

\subsubsection{Workers' Specialty}

% Workers' specialty level ${SL_{i}}$  is an indicator for a worker's preferences for specific task type ($WType$). Therefore, Workers' specialty level ${SL_{i}}$ is defined as the ratio of the frequency of worker $i$ registered for task type $k$ , (${WType_{k}}$)to all the different task types that worker $i$ registered for.

Workers' specialty level ${SL_{i,r}}$  is an indicator for worker i's preferences for a specific task type $r$. Therefore, workers' specialty level ${SL_{i,r}}$ is defined as the ratio of the frequency of worker $i$ registration for the task type $r$ , (${WType_{i,r}}$) to the frequency of all the different task types, ${j}$, that worker $i$ registered for (${WType_{i,j}}$).

\begin{equation}
{SL_{i,r}} = \frac{{WType_{i,r}}} 
{\sum_{j=0}^{n}{WType_{i,j}}}
\end{equation}

%%%%%%%%%%%%%%%%%%%%%%%%%%%%%%%%%%%%%%%

\begin{algorithm}
\caption{Collaborative Filtering Algorithm}
\label{CFA}
\SetAlgoLined
\indent
\DontPrintSemicolon
\begin{flushleft}
$I$ = set of all workers\; 
${WC_i}$ = set of collaborators to worker $i$ \;
${WT_i}$ = set of tasks registered by worker $i$ \; 
${Recom_{i}}$ = set of tasks recommended to worker $i$\; 
${poten_{ki}}$ = set of tasks registered by worker $k$ and not registered by worker $i$\; 
${cond}$ = set of conditions in eq()
\end{flushleft}
\For{ i $\in$ I}{
${WC_i}$ $=$ $\varnothing$\;
\For{j $\in$ (I/\{i\})}{
\If{${WT_i}$ $\cap$ ${ WT_j}$ $\neq$ $\varnothing$}{
${WC_i}$ = ${WC_i}$ $\cup$ $ j$\;
}
}
${Recom_{i}}$ = $\varnothing$\; 
\For{k $\in$ ${WC_i}$}{
${poten_{ki} = WT_k - WT_i}$\; 
\For{r $\in$ ${poten_{ki}}$}{
${cnt = 0}$\; 
\For{condition $\in$ ${\{condition_1, condition_2, condition_3\}}$}{
\If{$condition == TRUE$}{
cnt = cnt + 1\;
}
}
\If{cnt $\ge$ 2 $\And$ condition == TRUE}{
${Recom_{i}}$ $=$ ${Recom_{i}}$ $\cup$ $ r$\;
${SL\_set_i}$ $=$ ${SL\_set_i}$ $\cup$ $ SL_{i,r}$ \; 
$APL\_set_{i}$ = $APL\_set_{i}$ $\cup$ $APL_{i,r}$\;
}
}
}
${MX\_SL_i}$ $=$ $max(SL\_set_i)$ \; 
${MX\_APL_i}$ $=$ $max(APL\_set_i)$ \;
\For{${r2}$ $\in$ ${Recom_i}$}{
Proficiency = $APL_{i,r2}$ / $MX\_APL_i$ $\And$\;
Specialty = $ SL_{i,r2}$ / $MX\_SL_i$ \;
\If{Proficiency > 0.5 \& Specialty > 0.5}{
$Label_{i,r2}$ = ${very strong recommend}$\;
}
\If{Proficiency > 0.5 \& Specialty < 0.5}{
$Label_{i,r2}$ = ${ strong recommend}$ \; 
}
\If{Proficiency < 0.5 \& Specialty > 0.5}{
$Label_{i,r2}$ = ${recommend}$ \; 
}
\If{Proficiency < 0.5 \& Specialty < 0.5}{
$Label_{i,r2}$ = ${ low recommend}$ \; 
}
}
}
\end{algorithm}

%%%%%%%%%%%%%%%%%%%%%%%%%%%%%%%%%%%%%%

\subsection{Collaborative Recommendation} \label{colab}

Collaborative filtering recommendation \cite{terveen2001beyond} uses similarities between workers' profile and the history of their performances to provide serendipitous recommendations. To create the workers' collaborative filter, we use the workers collaboration as an input, then we add workers proficiency and specialty level, workers monetary prize and duration to narrow down the match between the recommended tasks and workers suitability.

In details as it is shown in Algorithm \ref{CFA}, system searches among the list of registrants of the tasks which was registered by worker $i$ (${W_{j}}$), if worker $j$ (${W_{j}}$) was registered for at least one of the tasks in the list then worker $j$ is added to the set of worker $i$'s collaborators, (${WC_{i}}$). Automatically, all the tasks that were registered by  worker $j$ and not worker $i$ is added to the list of potential tasks (${Poten_{k,i}}$) for worker $i$. Then system analyzes the potential tasks  based on worker$i$'s preferences in terms of monetary prize, task duration, proficiency (${{APL_{i}}}$), and specialty (${{SL_{i,r}}}$). Task $z$, ${t_{z}}$, should meet 3 out of 4 below conditions in order to be added to list of ${worker i}$'s recommended tasks, (${Recom_{i}}$):

\begin{itemize}
% \begin{flushleft}

    \item Condition 1.  ${{bMP_{i}} <= {MP_{z}}}$;
    \item Condition 2.  ${{bD_{i}} <= {D_{z}}}$ ;
    \item Condition 3. ${{APL_{i,r}} > \alpha}$;
    \item Condition 4.  ${{SL_{i,r}} > \beta}$;

% \end{flushleft}    
\end{itemize}

$\alpha$ and $\beta$ represent thresholds compared to which the minimum values of the proficiency (${APL_{i,r}}$) and specialty (${SL_{i,r}}$) should be higher. These information is chosen by the worker based on the characteristics of the task the worker has registered for. In this research we set both $\alpha$ and $\beta$ to 30\%. 
The recommended tasks will be mapped to two sets of values for specialty level (${{SL-set_{i}}}$) and average proficiency level (${{APL-set_{i}}}$) as an interval between 0 to 1. In next steps systems used the mapped data provides set of recommended tasks recommendation with four labels of \textit{"very high recommend"}, \textit{"high recommend"}, \textit{"recommend"}, and \textit{"week recommend"} per each active worker.

% \hl{Correspond to each qualified worker, for receiving a recommended task, there is a set of recommended set,${{Recom_{i}}}$, a set of values for ${{SL_{i,r}}}$ as ${{SL-set_{i}}}$, and a set of values for ${{APL_{i}}}$ as ${{APL-set_{i}}}$. To make the values in the sets more tangible, they are mapped into an interval between 0 and 1, dividing the all the set members by the maximum value in the set. Then the mapped values are used to categorize the recommended tasks to each worker into 4 group.}

% Collaborative filtering recommendation \cite{terveen2001beyond} uses similarities between workers' profile and their history of performance to provide recommendations. This allows for serendipitous recommendations.
% To create the workers' collaborative filter, we used the workers collaboration metric as an input, then we added workers proficiency and specialty level to narrow down the match between the recommended tasks and workers suitability. Algorithm \ref{CFA}  presents the associated pseudo code with the workers collaborative filtering recommendation.

% The result of this algorithm provides a set of recommended tasks, corresponding to each qualified worker. After the success predictor procedure, the tasks will be labeled as:  \textit{"very high recommend"}, \textit{"high recommend"}, \textit{"recommend"}, and \textit{"week recommend"}.

\begin{itemize}

    \item \textit{Very strong recommend}: is the list of recommended task that worker's proficiency and specialty level are greater than 50\%;
    
    \item \textit{Strong recommend}: is the list of task that workers has the specialty level less than 50\% but the proficiency level is greater than 50\%;

    \item \textit{Recommend}: is the list of task that workers has the specialty level greater than 50\% but the proficiency level is less than 50\%;

    \item \textit{Low recommend}: is the list of recommended task that worker's proficiency and specialty level are less than 50\%.
    
\end{itemize}

\subsection{Success Predictor}

Making a submission by itself does not guarantee workers' success, in order to be successful in a CSD platform, it is important to make a valid submission. However, making a submission by higher ranked workers may impact on workers decision on attempting to submit\cite{archak2010money}\cite{saremi2017leveraging}. It is important for a worker to be able to evaluate competitor's probability and trustworthiness in making a valid submission and as the result predict their own success.

\subsubsection{Specialty Participation Ratio}
Specialty Participation, ${{SP_{i}}}$, represents worker $i$’s registration frequency in tasks with the same type as the recommended task  $z$. Therefore, specialty participation ratio, ${{SPR_{i}}}$, defines as the ratio of specialty participation per worker, ${{sP_{i}}}$, to the total registration frequency in tasks with a similar  type by all the workers to whom the task $z$ is recommended to ${\sum_{j=0}^{N_{w}}{TP_{j}}}$. 

\begin{equation}
{SPR_{i}} = \frac{SP_{i}} 
{\sum_{j=1}^{N_{w}}{SP_{j}}} 
\end{equation}

\subsubsection{Average Proficiency Experience Ratio  ${{APER_{i}}}$}
 Proficiency experience ratio of worker $i$, ${{PER_{i,k}}}$, represents the level of proficiency in technology requirement $k$ to perform the recommended task $z$ in compare with the list of opponents who compete on the task $z$. It defines as the ratio of  the frequency which the worker $i$ has utilized the technology $k$ to the total utilization of technology$k$ by the workers to whom task $z$ is recommended to, ($N$). 
 
 \begin{equation}
{PER_{i,k}} = \frac{PE_{i,k}} 
{\sum_{j=1}^{N_{w}}{PE_{j,k}}} 
\end{equation}

In the case that the recommended task $z$ required multiple technologies,  the average proficiency experience ratio ${{APER_{i}}}$ will be used.

 \begin{equation}
{APER_{i}} = \frac{\sum_{j=1}^{N_{tsk}}{PER_{j,k}}} 
{N_{tsk}} 
\end{equation}

\subsubsection{Workers' Trustworthiness}

To analyze workers' trustworthiness we expand topcoder definition of reliability. In topCoder, crowd worker’s reliability of competing on the tasks is measured based on last 15 competitions workers registered and submitted. For example, if a worker submitted 14 tasks out of 15 last registered tasks, his reliability is 93\% (14/15).
% \hl{ Therefore, worker trustworthiness level, ${TL_{i}}$, in this research, measures the sum of valid submission, ${VS_{i}}$, to up to the last 15 registered tasks ,${R15_{i}}$, for worker $i$.}

Therefore, worker trustworthiness level, ${TL_{i}}$, in this research measures ratio of number of valid submissions, ${VS_{i}}$, for worker $i$ to register tasks up to the last 15 registered tasks ${R15_{i}}$.

\begin{equation}
{TL_{i}} = \frac{\sum_{i=1}^{15}{VS_{i}}} 
{\sum_{i=1}^{15}{R15_{i}}} 
\end{equation}

\subsubsection{Probability of  Valid Submission per Pelt}
Probability of valid submission per belt, ${p(VS)_{m}}$, measures the probability of worker $i$ from belt $b$ makes a valid submission.  

\begin{equation}
{p(VS)_{b}} = \frac{\sum_{b=1}^{n}{VS_{b}}} 
{\sum_{b=1}^{n}{R_{b}}}
\end{equation}

\subsubsection{Probability of Success}
Probability of success for worker $i$, ${p(Su)_{i}}$, who registers for a task is a function of the worker's specialty participation ratio ${{SPR_{i}}}$, average proficiency experience ratio ${{APER_{i}}}$, Trustworthiness ${TL_{i}}$, and  probability of  valid submission  ${p(VS)_{i}}$ , as below:

\begin{equation}
{p(Su)_{i}} = \frac{{SPR_{i}}*{APER_{i}}*{TL_{i}}*{p(VS)_{i}}} 
{\sum_{i=1}^{N_{w}}{{SPR_{j}}*{APER_{j}}*{TL_{j}}*{p(VS)_{j}}}}
\end{equation}

\section{Experiment Design}

To evaluate the conceptual model introduced in Section \ref{model}, this section presents the experiment design and evaluation base line for this study.

\subsection{Research Questions}

To investigate the  impact of workers' collaboration, proficiency and specialty in workers' success, the following research questions were formulated and studied in this paper:

\textit{RQ1 (Overall worker Performance)}: How distributed are crowd
workers in terms of expertise?

This research question aims at providing general overview of workers distribution in the platform  based on their proficiency and specialty in performing tasks;

\textit{RQ2 (Worker recommendation)}: How  to strategically take a new task to ensure workers' success?

Understanding opponents performance pattern in a competition  helps to provide better decision making for taking a task.

\subsection{Dataset}  \label{data}

The dataset from topcoder contains 403 individual projects including 4,907 component development tasks (ended up with 4,770 after removing tasks with incomplete information) and 8,108 workers from January 2014 to February 2015 (13 months). Tasks are uploaded as competitions in the platform, where crowd software workers would register and complete the challenges. When the workers submit the final files, it will be reviewed by experts to check the results and grant the scores.

The dataset contains tasks attributes such as  required technology, platform, task description, task status, monetary prize, days to submit, registration date, submission date, and workers attributes such as registration date, submission date, valid submission, winning placement, winning status, rating score, and winning score. In this step, workers' rank, proficiency, specialty and trustworthiness metrics are not included. Then, we create attributes, such as workers skillsets (WTech), workers task type (WType), 
% and workers' collaboration (WC)  
which are proxy by the  number of technologies (\#Tech) required to perform taken tasks by the worker, and different task type worker can choose to work on. 
% and the history of two workers competing on the same task.
We create binary variables for each technology (WTech) required in each task, and each task type (WType) worker chose to perform,
% and workers' collaboration (WC),
where:

\[
WTech(x,p) =
	\begin{cases}
        1 &\parbox[t]{.35\textwidth}{worker x is an expert in technology p} \\
        0 & \text{otherwise}
    \end{cases}
\]

and, 

\[
WType(x,s) =
	\begin{cases}
        1 &\parbox[t]{.35\textwidth}{worker x is a specialist in task type s} \\
        0 & \text{otherwise}
    \end{cases}
\]
% and,
% \[
% {WC_{i,j}} =
% 	\begin{cases}
%         1 &\parbox[t]{.35\textwidth}{${W_{i}}$ and ${W_{j}}$ registered for the same task} \\
%         0 & \text{otherwise}
%     \end{cases}
% \]

The worker attributes used in the analysis are presented in top section of Table \ref{metrics}.

\subsection{Implementation of The Collaborative Recommendation System }

There are six steps in implementing the collaborative recommendation system: workers' collaboration, workers' monetary prize, workers' duration, workers' proficiency, workers' specialty and workers' success predictor.

\subsubsection{Workers' Collaboration:}
Workers' Collaboration is analyzed based on the  bipartite network of workers. If two workers registered on the same tasks, they are collaborating. In this research, first we grouped workers’ attribute per month and labeled workers who had minimum one registration per month as active worker. This reduced the number of workers from 8180 to 2259 workers. Then we analyzed workers' collaboration based on the entire dataset (i.e 13 months from Jan 2014 to Feb 2015) explained in part \ref{data}.

\subsubsection{Workers' Monetary Prizes:}
Worker's monetary prize is analyzed based on the minimum monetary prize in the history of tasks worker $i$ registered for.
% For example, if worker $i$ was competing on tasks with minimum monetary prize of 500\$, then system recommend tasks with no associated prize less than 500\$.

for example, worker I in part \ref{example}has the history of performing on tasks with the minimum monetary prize of 750\$.  Among available tasks that worker I can potentially take,  tasks 1 and 3 are offering higher monetary prize than worker's monetary prize (i.e 800\$ and 2500\$ respectively) and can be labeled as potential recommended task.

\subsubsection{Workers' Duration:}
In this research, duration among the tasks worker $i$ compete on is used as the base duration for recommendation. For instance 
% if the minimum duration if tasks worker $i$  registered for is 7 days system will recommend him/her a task with no shorter than 7 days duration. 
worker I in part \ref{example} has the history of taking tasks with duration as short as 7 days.  Therefore, in terms of duration, all the 3 tasks meet worker I's preference, figure\ref{motivating}, and system will labeled them as potential recommended task.

\subsubsection{Workers' Proficiency:}
Workers' proficiency is designed to analyze the match between workers' skill set and tasks requirement per technology. For example, if task ($l$), $ {T_{l}} $, listed \textit{HTML} as one of the requirements and worker(i), $ {W_{i}} $, compete on ${a}$ task/s with similar requirements out of total ${b}$ tasks in workers' history, then $ {W_{i}} $ has ${a/b}$ proficiency level to perform $ {T_{l}} $.

In this research the workers' proficiency calculated based on binary metrics ${Wtech_{i}}$ and  provide the percentage of match between workers and potential suggested tasks. 
According table \ref{workers} to register for task 1, Worker I has proficiency level of 0.307(i.e (100/(100+67+60+67+32))) in Java.

\subsubsection{Workers' Specialty:}
Workers' Specialty measures the level of the match between a task and workers historical performance on similar task type. For example, if task(l), $ {T_{l}} $, labeled as type \textit{code} in competitions and worker(i), $ {W_{i}} $, compete on ${c}$ task/s with similar task type out of total ${b}$ tasks in workers' history, then $ {W_{i}} $ has ${c/b}$ Specialty level to perform $ {T_{l}} $.

To analyze the specialty match this research used the introduced metric ${WType_{i}}$ in part \ref{data}. For example, based on table \ref{workers} to register for task 1, Worker I has specialty level of 0.30 (i.e 100/(100+42)) to perform First2Finish task type.

\subsubsection{Workers' Success Predictor:}
To predict workers success in taking the recommended tasks , this research analyzed workers probability of success based on the registered opponent probability of making a valid submission for the task.

\begin{figure}[ht!]
\centering
\includegraphics[width=1\columnwidth]{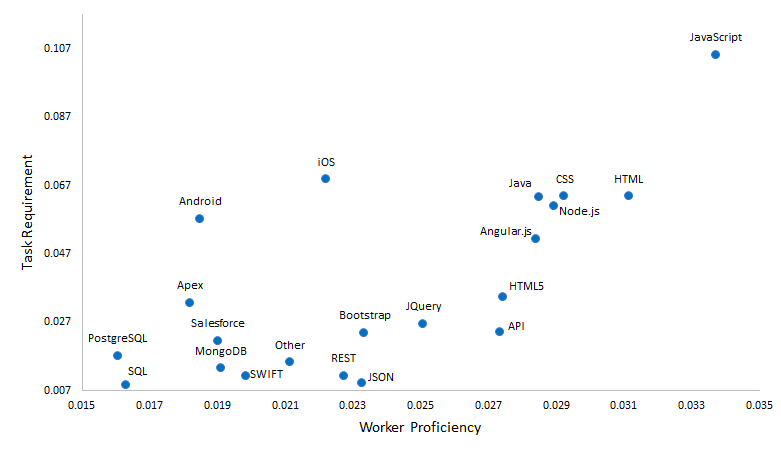}
\caption{Overall Distribution of Workers' Proficiency per Technology}
\label{prof}
\end{figure}

\section{Result and Discussion}

\begin{table*}[!ht]
\caption{Summary of  Workers' Profile Overview in Motivating Example} % title of Table
\centering % used for centering table
\begin{tabular}{p{1.5cm} p{1.5cm} p{1.5cm} p{1cm}p{1cm} p{1.5cm}p{1.25cm} p{3cm} p{3cm}}
\hline
Worker ID  & \# R & \# VS & \# Collaborators  & Belt & Min(MP) & Min(D) & Top 5 Proficiency & Top 2 Specialty\\ %[0.5ex] % inserts table
\hline%\hline % inserts single horizontal line
Worker I & 244 & 80 & 61 & Green & 750 & 7 & Java(100), SQL(67), Android(60), Apex(67), .NET(32)
& UI Prototype Competition (100), First2Finish(42) \\
Worker II & 150 & 87 & 80 & Gray  & 890 & 23& Android (60), MongoDB (40), Node.js(32), JavaScript(30), Java(16)
&  First2Finish(96), Code(80)\\
Worker III & 883 & 32 & 69 & Yellow & 1000 & 10 & Java(300), IOS(210), MOngoDB(163), Angular.js(80), Node.js(65)
 &  Assembly Competition (300),  First2Finish (180)
\\
Worker IV & 195 & 14& 85 & Blue & 1250 & 16 & Java(70), IOS(56), MOngoDB(50), .NET (45), Node.js(45)
 & Assembly Competition (78),  UI Prototype Competition (60)
\\
  
\hline
\label{workers}
\end{tabular}
\end{table*}

\subsection{Overall Worker Performance}

In order to have a better understanding of workers' overall performance, we studied overall distribution of workers' proficiency and workers' specialty based on required technology by tasks and tasks type in the platform. 

\subsubsection{Workers' Proficiency:} We identified 115 technologies in the dataset. Mapping among technologies, task and workers showed that 50\% of tasks can be done by knowledge of only 8 technologies, and 50\% or workers are actively taking tasks with requirements of one of the top 21 technologies. Therefor in  this part we visualize the workers' proficiency for the top 21 technologies in the dataset. 
According to figure \ref{prof}, the most on demand technology is JavaScript with 3.4\% of workers' proficiency and almost 11\% of available tasks to compete on, the second technology is HTML with 3.1\% workers' proficiency and 6\% of tasks as requirements. CCS is the third technology that contains 3\% workers proficiency. Interestingly, the technology with the least level of proficiency is PostgreSQL followed by SQL with almost 1.6\% workers' proficiency.
The rest of technologies attract workers with on average 2.3\% proficiency and provide a pool of task with requirement technology between 3.2\% to 6\%. 

\begin{figure}[ht!]
\centering
\includegraphics[width=1\columnwidth]{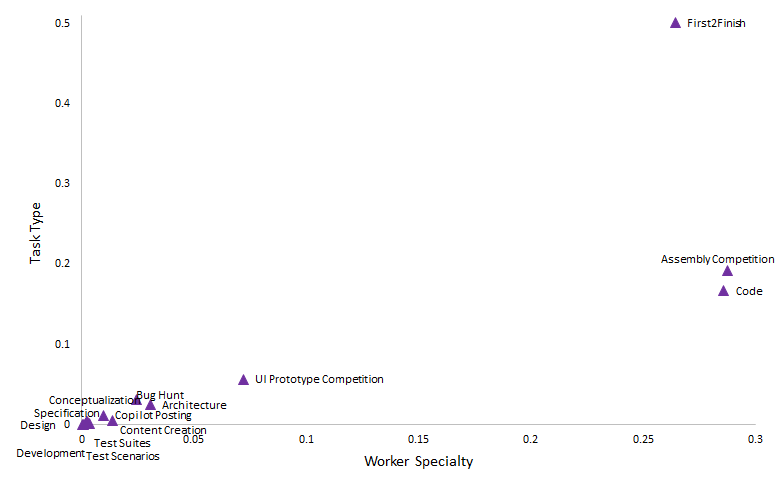}
\caption{Overall Distribution of Workers' Specialty per Task Type}
\label{splt}
\end{figure}

\subsubsection{Workers' Specialty:} Topcoder introduced 14 task type to compete on. Mapping among technologies, task and workers showed that 50\% of workers are interested in Assembly  Competitions, assembling previous tasks, and Code, Programming specific task,  while 50\% of tasks are under First2Finish type, in which The first person to submit passing entry wins.

As it is illustrated in figure \ref{splt}, workers' specialty for First2Finish task type is 26\%, while for  Assembly and Code is 29\% and 28\% respectively. The next task type is UI Prototype Competition with 7\% workers' specialty and almost 6\% task in the platform.  The rest of tasks types are attracting workers with less than 3\% specialty.

\subsection{Worker Recommendation}

We trained the proposed recommendation system based on 11 months data from our data set s (from Jan 2014 to Jan 2015). Then we recommended task to active workers during Jan 14\textsuperscript{th} 2015 to Jan 30\textsuperscript{th} 2015. 
% This section presents the evaluation results of applying the proposed recommendation system for active crowd workers during Jan 14\textsuperscript{th} 2015 to Jan 30\textsuperscript{th} 2015. 
According to our dataset there are 260 active workers and 311 available tasks in the platform during this period.
The initial result of the recommendation system provide at least 1 recommended tasks for 237 workers, and no suitable tasks for 23 workers. Also workers would succeed with average probability of 23\%. The maximum probability of success was 86\% and minimum was 2\%. 165 workers(i.e 57\%) would receive success will probability of success less than average. 

In order to understand how the system works, we picked workers from part \ref{example} and explain all the steps system takes. As it is summarized in table \ref{workers}, \textit{worker I} is from Green belt, competed on 244 tasks with 61 collaborators. \textit{Worker II} belongs to  gray rank, with 150 competition in the profile and 80 collaborators. \textit{Worker III} is one of high ranked workers from Yellow community, who competed 883 times with 69 collaborators. and finally \textit{worker IV} is a blue belt with 195 competition and 85 collaborators.

\subsubsection{Collaborative Recommendation}
As it will be explained in details in this part, recommendation system makes recommendation from 19 tasks to 4 chosen worker to register for next 14 days. Each workers received a list of recommended task between 11 to 18 task. Each task is labeled as one of the recommendation levels introduced in part \ref{colab}. 

\begin{figure}[ht!]
\centering
\includegraphics[width=0.9\columnwidth]{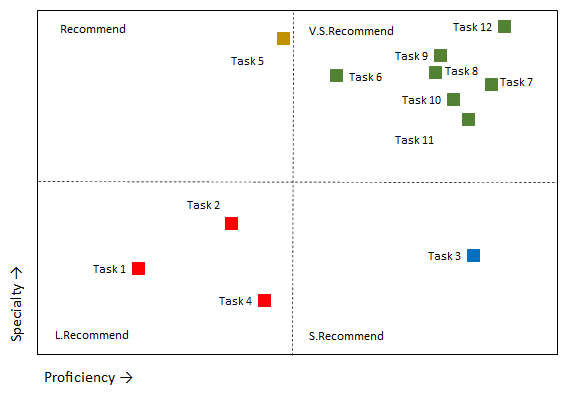}
\caption{Magic Quadrant for Worker I Recommended Tasks}
\label{I}
\end{figure}

\textit{ Worker I:} As figure \ref{I} represents, the recommendation system recommends 12 tasks to worker I to take (i.e tasks 1,2,3,4,5,6,7,8,9,10,11,12).
Task 1, 2 and 4 are under low recommended section, which mean worker I's proficiency and specialty in performing this tasks is under 50\%. Task 5 is located in the recommended section. This means however worker I does not have high level of proficiency to perform this tasks but s/he has is specialized to do so. Task 3 is in strong recommended which relates to high level of proficiency of worker I to perform this task, there are 7 tasks (6,7,8,9,10,11 and 12) that are very strongly recommended to be taken by worker I.

\begin{figure}[ht!]
\centering
\includegraphics[width=0.9\columnwidth]{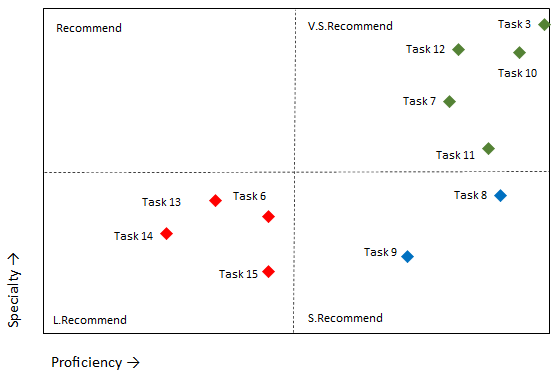}
\caption{Magic Quadrant for Worker II Recommended Tasks}
\label{II}
\end{figure}

\textit{ Worker II:} As figure \ref{II} presents, Worker II is recommended to register for 11 tasks (tasks 3,6,7,8,9,10,11,12,13,14,15).
From these 11 tasks, tasks 6,13,14 and 15 are labeled as low recommended. Worker II does not received any recommended task under recommended section. S/he has tasks 8 and 9 under strong recommended  with high proficiency and low specialty. And s/he has 5 suggested tasks in the very strong recommended section to take for next tow weeks(i.e tasks 3,7,10,11 and 12).

\begin{figure}[ht!]
\centering
\includegraphics[width=0.9\columnwidth]{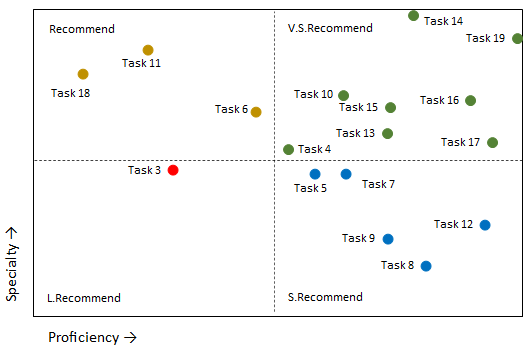}
\caption{Magic Quadrant for Worker III Recommended Tasks}
\label{III}
\end{figure}

\textit{ Worker III:} Figure \ref{III} illustrates the magic quadrant for worker III's recommended tasks. According to figure \ref{III} , 17 tasks are recommended to worker III (tasks 3,4,5,6,7,8,9,10,11,12,13,14,15,16,17,18,19).
 4 out of 17 tasks are under low recommended (task 3) or recommended (tasks 11,16,18) section. 
 Tasks 5,7,9,8,and 12 are located under the strong recommended tasks and tasks 4, 10, 13, 14, 15, 16, 17 and 19 are very strongly recommended to worker III to take them based on her/his proficiency and specialty level. 

\begin{figure}[ht!]
\centering
\includegraphics[width=0.9\columnwidth]{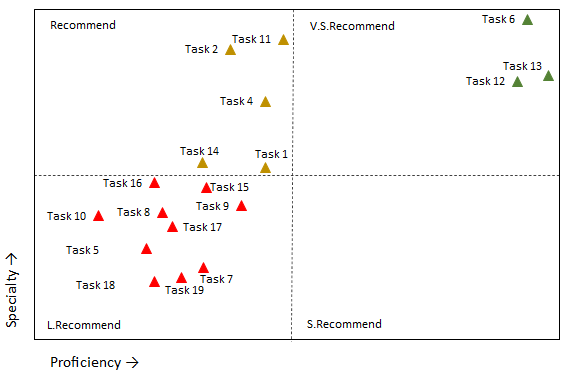}
\caption{Magic Quadrant for Worker IV Recommended Tasks}
\label{IV}
\end{figure}

\textit{ Worker IV:} As it is clear in figure \ref{IV} The recommendation system provide 18 tasks to worker IV to potentially register for, tasks 1,2,4,5,6,7,8,9,10,11,12,13,14,15,16,17,18 and 19.
4 out of 17 tasks are under low recommended (task 3) or recommended (tasks 11,16,18) section. While 15 out of 18 tasks are labeled under recommended or low recommended, there is no task under strong recommended and only 3 tasks of 6, 12 and 13 are very strongly recommended to be taken by worker IV. 

\subsection{Success predictor}
It is not practical to register and compete on minimum 12 tasks during two weeks, therefore, system calculate the probability of success per recommended tasks for the workers based on registered opponents to compete on tasks and suggest the top three tasks to workers. Table \ref{Wsuccess} presents the order of tasks per worker.

\begin{table}[!ht]
\caption{Final Task Suggestion to Workers} % title of Table
\centering % used for centering table
\begin{tabular}{p{2cm} p{1cm} p{2cm} p{1cm}}
\hline
\\
Worker ID & Task Order & Recommended Level &  p(success)\\ %[0.5ex] % inserts table

\hline%\hline % inserts single horizontal line

\multirow{4}{*}{Worker I} 
\\
& Task 12 &  V.S.R & 30\%  \\
& Task 8 &  V.S.R & 15\%  \\
& Task 9 &  V.S.R & 15\%  \\
\\
\hline

\multirow{4}{*}{Worker II} 
\\
& Task 8 &  S.R & 43\%  \\
& Task 9 &  S.R & 40\%  \\
& Task 10 &  V.S.R & 39\%  \\
\\
\hline
\\
\multirow{4}{*}{Worker III} 
\\
& Task 17 &  V.S.R & 58\%  \\
& Task 16 &  V.S.R & 56\%  \\
& Task 18 &  R & 50\%  \\
\\
\hline

\multirow{4}{*}{Worker IV} 
\\
& Task 6 &  V.S.R & 17\%  \\
& Task 13 &  V.S.R & 16\%  \\
& Task 2 &  R & 10\%  \\
\\
\hline

\label{Wsuccess}
\end{tabular}
\end{table}

As it is summarized in table\ref{Wsuccess} worker I has the highest chance of success in taking task 12 among all the list of recommended tasks with 30\% probability of success, the second highest chance happens when worker I takes either task 8 or task 9, both tasks provide 15\% probability of success for worker I.
For worker II task 8 provides the highest probability of success of 43\%, Task 9 provides 40\% probability of success and task 10 brings 39\% probability of success. Interestingly, the top two highest probability of success for worker II come with tasks from strong recommended group, meaning worker II has high proficiency in technical term to perform the tasks, while the specialty in task type is not very high (i.e. under 50\%). 
Worker III has 58\% chance of success with registering for task 17 followed by 56\% probability of success for task 16. both tasks are under very strongly recommended group. The probability of success drops to 50\% for task 18 as the third suggested task from recommended cluster.
The last worker (i.e worker IV) will have the highest probability of success by competing on task 6 for17\% chance, followed by 16\% probability of success for task 13. and the third task with highest probability of success is task 2 with 10\% probability of success which is under recommended group of tasks for worker IV.

% Figure \ref{success}-a. represents the order of recommended tasks for worker I. Worker I has the highest probability of success  of 30\% in taking tasks 12, followed by 15\% probability of success in tasks 8 and 9. Tasks 7 provides 7\% success probability and taking rest of the tasks bring 5\% chance in success. 
% Worker II have opportunity to have success more than 30\% in 5 tasks, started with 43\% in taking task 8 Figure \ref{success}-b. 
% Worker III Figure \ref{success}-c. would have more than 55\% chance of success by taking either tasks 17,16,18,or 19. however tasks 17 provides the highest level of success with 58\% probability of success. Tasks 16,18 and 19 would provides the same level of success, however tasks 16 and 19 are in the list of very strongly recommended tasks while task 18 is under recommended tasks. This fact may impact on worker III decision making since s/he does not have high proficiency level in performing tasks 18.   
% As for worker IV, tasks  6 and 13 will lead to more than 10\% chance of success ( 17\% and 16\%  respectively) and the rest of recommended tasks will provide probability of less than 10\% success.

% \begin{figure*}[ht!]
% \centering
% \includegraphics[width=0.9\textwidth]{success.png}
% \caption{Order of Recommended Tasks based on Probability of Success}
% \label{success}
% \end{figure*}

% \begin{figure*}[ht!]
% \centering
% \includegraphics[width=0.9\textwidth]{order.png}
% \caption{Order of Recommended Tasks based on Probability of Success}
% \label{success}
% \end{figure*}

\subsection{Validation} 

The last step was to understand the accuracy of the recommendation system for different workers. To do so, the mean relative error (${MRE}$) of each workers' list of recommended task  (${Recom_{i}}$) based on the available actual workers' task registration (${R_{i}}$) in the same period of time (Jan 14\textsuperscript{th} 2015 to Jan 30\textsuperscript{th} 2015) was calculated, where ${y}$ is number recommended task to worker${i}$, and ${n}$ is number of actual tasks that worker ${i}$ registered for as displayed below.

\[
{MRE_{i}} = \frac{\sum_{i =0}^{n}{R_{i}}- \sum_{i=0}^{y}{Recom_{i}}} {\sum_{i =0}^{n}{R_{i}}}
\]

The t-test was also applied to the prediction results in each state to confirm the models’ accuracy.

The mean relative error (${MRE}$) of the recommendation system was only 1.9\% . The results of the t-test on 311 tasks that recommended to 260 workers revealed that the probability of error in recommendation system was 0.012  with zero hypothesized mean difference. This result shows that the proposed recommendation system is performing accurately.

\subsection{Discussion} 
Recommending a task is dynamic in nature since new tasks are uploaded all the time, and some other tasks are getting finished. Consequently, a proper recommendation needs to be regularly updated. Different factors impact a worker performance in CSD, to accurately support workers dynamic decision making process it is important to understand and capture the variation of these factors.  The proposed collaborative recommendation system in this paper addresses the measurement of varying amount of proficiency and specialty with respect to required technology and task type
and preferences of individual workers using monetary prize and duration.
From workers perspectives, success-driven
task recommendation systems helps to: 1)  ensures success and increasing expertise of workers in short time, 
and 2) guides the healthy growth of
competition for tasks with respect to the new and emerging technologies.
Even though crowd workers try to increase their proficiency level in new technologies, they may not pay enough attention to
demanding technologies \cite{karim2018learn}.

\subsection{Threats to Validity}

First, the study only focuses on competitive CSD tasks on the topcoder platform. Many more platforms do exist, and even though the results achieved are based on a comprehensive set of about 60,000 task-worker, the results cannot be claimed externally valid. There is no guarantee the same results would remain exactly the same in other CSD platforms.

Second, there are many different factors that may influence workers' preference, performance and decision in task selection and completion. Our worker collaboration approach is based on known task-worker attributes in topcoder. Different approaches may lead us to different but almost similar results.

Third, the result is based on network of tasks-workers only. Workers communication was not considered in this research. In future we need to add this level of research to the existing one.

\subsection{Adaptability to Different Platforms}
The overall presented collaborative recommendation system in this paper is adoptable to different crowdsourced platforms. However, based on the type of platform ( i.e competitive, collaborative or coopetitive) the workers success predictor needs to be updated by different task validation and workers reliability algorithm adopted by the chosen platform.
To make the presented system compatible for different crowdsourcing platforms, there is no need to update any part of the collaborative recommended system but the back end analysis in elaborating defined input metrics.

\section{Conclusion and future work}

Crowdsourced Software Development (CSD) is an
emerging paradigm that has been increasingly adopted.
In a competitive crowdsourcing marketplace, competition for success over shared demand adds  uncertainty in decision making process for crowd workers. Therefor, it is critical for a crowd worker to not only understand the suitability of available tasks to take, but also the sensitivity and performance of the opponents in taking the tasks and rate of  success.
This paper reports a collaborative recommendation system for workers  to address that end.

The proposed collaborative recommendation system nominates tasks with high suitability to workers  based on a set of workers attributes such as workers' proficiency and specialty in performing the task, workers' trustworthiness in make a valid submission, minimum monetary prize and minimum task duration, and workers' collaboration with other opponents. Then the system provide probability of success per recommended task based on other workers who received the task in their recommendation list( potential opponents).

The proposed collaborative recommendation system empowers crowd workers to explore different potential success strategies with respect to different available tasks and active opponents in the platform. This includes the probability of workers success, task duration, monetary prize, task type and required technology.
Moreover, experimental results on Experimental results on 260 active crowd workers demonstrate that just following the top three success probability of task recommendations, workers can achieve success up to 86\%. 

In future, we would like to expand our system and add workers communication and relation to the system which develops
facilitating techniques to support decisions during task
scheduling phase.

% \bibliographystyle{ACM-Reference-Format}
\bibliographystyle{ACM-Reference-Format}
\bibliography{bibfile.bib}


\begin{thebibliography}{40}

%%% the Web address.

\ifx \showCODEN    \undefined \def \showCODEN     #1{\unskip}     \fi
\ifx \showDOI      \undefined \def \showDOI       #1{#1}\fi
\ifx \showISBNx    \undefined \def \showISBNx     #1{\unskip}     \fi
\ifx \showISBNxiii \undefined \def \showISBNxiii  #1{\unskip}     \fi
\ifx \showISSN     \undefined \def \showISSN      #1{\unskip}     \fi
\ifx \showLCCN     \undefined \def \showLCCN      #1{\unskip}     \fi
\ifx \shownote     \undefined \def \shownote      #1{#1}          \fi
\ifx \showarticletitle \undefined \def \showarticletitle #1{#1}   \fi
\ifx \showURL      \undefined \def \showURL       {\relax}        \fi
% invisible to TeX
\providecommand\bibfield[2]{#2}
\providecommand\bibinfo[2]{#2}
\providecommand\natexlab[1]{#1}
\providecommand\showeprint[2][]{arXiv:#2}

\bibitem[\protect\citeauthoryear{Abhinav, Bhatia, Dubey, Jain, and
  Bhardwaj}{Abhinav et~al\mbox{.}}{2020}]%
        {abhinav2020tasrec}
\bibfield{author}{\bibinfo{person}{Kumar Abhinav},
  \bibinfo{person}{Gurpriya~Kaur Bhatia}, \bibinfo{person}{Alpana Dubey},
  \bibinfo{person}{Sakshi Jain}, {and} \bibinfo{person}{Nitish Bhardwaj}.}
  \bibinfo{year}{2020}\natexlab{}.
\newblock \showarticletitle{TasRec: a framework for task recommendation in
  crowdsourcing}. In \bibinfo{booktitle}{\emph{Proceedings of the 15th
  International Conference on Global Software Engineering}}.
  \bibinfo{pages}{86--95}.
\newblock


\bibitem[\protect\citeauthoryear{Archak}{Archak}{2010}]%
        {archak2010money}
\bibfield{author}{\bibinfo{person}{Nikolay Archak}.}
  \bibinfo{year}{2010}\natexlab{}.
\newblock \showarticletitle{Money, glory and cheap talk: analyzing strategic
  behavior of contestants in simultaneous crowdsourcing contests on TopCoder.
  com}. In \bibinfo{booktitle}{\emph{Proceedings of the 19th international
  conference on World wide web}}. \bibinfo{pages}{21--30}.
\newblock


\bibitem[\protect\citeauthoryear{Archak and Ghose}{Archak and Ghose}{2010}]%
        {archak2010learning}
\bibfield{author}{\bibinfo{person}{Nikolay Archak} {and}
  \bibinfo{person}{Anindya Ghose}.} \bibinfo{year}{2010}\natexlab{}.
\newblock \showarticletitle{Learning-by-doing and project choice: a dynamic
  structural model of crowdsourcing}.
\newblock \bibinfo{journal}{\emph{LEARNING}}  \bibinfo{volume}{1}
  (\bibinfo{year}{2010}), \bibinfo{pages}{1--2010}.
\newblock


\bibitem[\protect\citeauthoryear{Bista, Nepal, Colineau, and Paris}{Bista
  et~al\mbox{.}}{2012}]%
        {bista2012using}
\bibfield{author}{\bibinfo{person}{Sanat~Kumar Bista}, \bibinfo{person}{Surya
  Nepal}, \bibinfo{person}{Nathalie Colineau}, {and} \bibinfo{person}{Cecile
  Paris}.} \bibinfo{year}{2012}\natexlab{}.
\newblock \showarticletitle{Using gamification in an online community}. In
  \bibinfo{booktitle}{\emph{8th International Conference on Collaborative
  Computing: Networking, Applications and Worksharing (CollaborateCom)}}. IEEE,
  \bibinfo{pages}{611--618}.
\newblock


\bibitem[\protect\citeauthoryear{Cavusoglu, Li, and Huang}{Cavusoglu
  et~al\mbox{.}}{2015}]%
        {cavusoglu2015can}
\bibfield{author}{\bibinfo{person}{Huseyin Cavusoglu}, \bibinfo{person}{Zhuolun
  Li}, {and} \bibinfo{person}{Ke-Wei Huang}.} \bibinfo{year}{2015}\natexlab{}.
\newblock \showarticletitle{Can gamification motivate voluntary contributions?
  The case of StackOverflow Q\&A community}. In
  \bibinfo{booktitle}{\emph{Proceedings of the 18th ACM conference companion on
  computer supported cooperative work \& social computing}}.
  \bibinfo{pages}{171--174}.
\newblock


\bibitem[\protect\citeauthoryear{Chandler and Kapelner}{Chandler and
  Kapelner}{2013}]%
        {CHANDLER2013123}
\bibfield{author}{\bibinfo{person}{Dana Chandler} {and} \bibinfo{person}{Adam
  Kapelner}.} \bibinfo{year}{2013}\natexlab{}.
\newblock \showarticletitle{Breaking monotony with meaning: Motivation in
  crowdsourcing markets}.
\newblock \bibinfo{journal}{\emph{Journal of Economic Behavior and
  Organization}}  \bibinfo{volume}{90} (\bibinfo{year}{2013}),
  \bibinfo{pages}{123--133}.
\newblock
\showISSN{0167-2681}
\urldef\tempurl%
\url{https://doi.org/10.1016/j.jebo.2013.03.003}
\showDOI{\tempurl}


\bibitem[\protect\citeauthoryear{Cheng and Zachry}{Cheng and Zachry}{2020}]%
        {cheng2020building}
\bibfield{author}{\bibinfo{person}{Ruijia Cheng} {and} \bibinfo{person}{Mark
  Zachry}.} \bibinfo{year}{2020}\natexlab{}.
\newblock \showarticletitle{Building Community Knowledge In Online
  Competitions: Motivation, Practices and Challenges}.
\newblock \bibinfo{journal}{\emph{Proceedings of the ACM on Human-Computer
  Interaction}} \bibinfo{volume}{4}, \bibinfo{number}{CSCW2}
  (\bibinfo{year}{2020}), \bibinfo{pages}{1--22}.
\newblock


\bibitem[\protect\citeauthoryear{Crump, McDonnell, and Gureckis}{Crump
  et~al\mbox{.}}{2013}]%
        {crump2013evaluating}
\bibfield{author}{\bibinfo{person}{Matthew~JC Crump}, \bibinfo{person}{John~V
  McDonnell}, {and} \bibinfo{person}{Todd~M Gureckis}.}
  \bibinfo{year}{2013}\natexlab{}.
\newblock \showarticletitle{Evaluating Amazon's Mechanical Turk as a tool for
  experimental behavioral research}.
\newblock \bibinfo{journal}{\emph{PloS one}} \bibinfo{volume}{8},
  \bibinfo{number}{3} (\bibinfo{year}{2013}).
\newblock


\bibitem[\protect\citeauthoryear{Difallah, Demartini, and
  Cudr{\'e}-Mauroux}{Difallah et~al\mbox{.}}{2013}]%
        {difallah2013pick}
\bibfield{author}{\bibinfo{person}{Djellel~Eddine Difallah},
  \bibinfo{person}{Gianluca Demartini}, {and} \bibinfo{person}{Philippe
  Cudr{\'e}-Mauroux}.} \bibinfo{year}{2013}\natexlab{}.
\newblock \showarticletitle{Pick-a-crowd: tell me what you like, and i'll tell
  you what to do}. In \bibinfo{booktitle}{\emph{Proceedings of the 22nd
  international conference on World Wide Web}}. \bibinfo{pages}{367--374}.
\newblock


\bibitem[\protect\citeauthoryear{Difallah, Demartini, and
  Cudr{\'e}-Mauroux}{Difallah et~al\mbox{.}}{2016}]%
        {difallah2016scheduling}
\bibfield{author}{\bibinfo{person}{Djellel~Eddine Difallah},
  \bibinfo{person}{Gianluca Demartini}, {and} \bibinfo{person}{Philippe
  Cudr{\'e}-Mauroux}.} \bibinfo{year}{2016}\natexlab{}.
\newblock \showarticletitle{Scheduling human intelligence tasks in multi-tenant
  crowd-powered systems}. In \bibinfo{booktitle}{\emph{Proceedings of the 25th
  international conference on World Wide Web}}. \bibinfo{pages}{855--865}.
\newblock


\bibitem[\protect\citeauthoryear{Faradani, Hartmann, and Ipeirotis}{Faradani
  et~al\mbox{.}}{2011}]%
        {faradani2011s}
\bibfield{author}{\bibinfo{person}{Siamak Faradani}, \bibinfo{person}{Bj{\"o}rn
  Hartmann}, {and} \bibinfo{person}{Panagiotis~G Ipeirotis}.}
  \bibinfo{year}{2011}\natexlab{}.
\newblock \showarticletitle{What’s the right price? pricing tasks for
  finishing on time}. In \bibinfo{booktitle}{\emph{Workshops at the
  Twenty-Fifth AAAI Conference on Artificial Intelligence}}.
\newblock


\bibitem[\protect\citeauthoryear{Farrell and Rabin}{Farrell and Rabin}{1996}]%
        {farrell1996cheap}
\bibfield{author}{\bibinfo{person}{Joseph Farrell} {and}
  \bibinfo{person}{Matthew Rabin}.} \bibinfo{year}{1996}\natexlab{}.
\newblock \showarticletitle{Cheap talk}.
\newblock \bibinfo{journal}{\emph{Journal of Economic perspectives}}
  \bibinfo{volume}{10}, \bibinfo{number}{3} (\bibinfo{year}{1996}),
  \bibinfo{pages}{103--118}.
\newblock


\bibitem[\protect\citeauthoryear{Gray, Suri, Ali, and Kulkarni}{Gray
  et~al\mbox{.}}{2016}]%
        {gray2016crowd}
\bibfield{author}{\bibinfo{person}{Mary~L Gray}, \bibinfo{person}{Siddharth
  Suri}, \bibinfo{person}{Syed~Shoaib Ali}, {and} \bibinfo{person}{Deepti
  Kulkarni}.} \bibinfo{year}{2016}\natexlab{}.
\newblock \showarticletitle{The crowd is a collaborative network}. In
  \bibinfo{booktitle}{\emph{Proceedings of the 19th ACM conference on
  computer-supported cooperative work \& social computing}}.
  \bibinfo{pages}{134--147}.
\newblock


\bibitem[\protect\citeauthoryear{Hettiachchi, Van~Berkel, Kostakos, and
  Goncalves}{Hettiachchi et~al\mbox{.}}{2020}]%
        {hettiachchi2020crowdcog}
\bibfield{author}{\bibinfo{person}{Danula Hettiachchi}, \bibinfo{person}{Niels
  Van~Berkel}, \bibinfo{person}{Vassilis Kostakos}, {and}
  \bibinfo{person}{Jorge Goncalves}.} \bibinfo{year}{2020}\natexlab{}.
\newblock \showarticletitle{CrowdCog: A Cognitive skill based system for
  heterogeneous task assignment and recommendation in crowdsourcing}.
\newblock \bibinfo{journal}{\emph{Proceedings of the ACM on Human-Computer
  Interaction}} \bibinfo{volume}{4}, \bibinfo{number}{CSCW2}
  (\bibinfo{year}{2020}), \bibinfo{pages}{1--22}.
\newblock


\bibitem[\protect\citeauthoryear{Howe}{Howe}{2008}]%
        {howe2008crowdsourcing}
\bibfield{author}{\bibinfo{person}{Jeff Howe}.}
  \bibinfo{year}{2008}\natexlab{}.
\newblock \bibinfo{booktitle}{\emph{Crowdsourcing: How the power of the crowd
  is driving the future of business}}.
\newblock \bibinfo{publisher}{Random House}.
\newblock


\bibitem[\protect\citeauthoryear{Hu and Wu}{Hu and Wu}{2014}]%
        {hu2014game}
\bibfield{author}{\bibinfo{person}{Zhenghui Hu} {and} \bibinfo{person}{Wenjun
  Wu}.} \bibinfo{year}{2014}\natexlab{}.
\newblock \showarticletitle{A game theoretic model of software crowdsourcing}.
  In \bibinfo{booktitle}{\emph{2014 IEEE 8th International Symposium on Service
  Oriented System Engineering}}. IEEE, \bibinfo{pages}{446--453}.
\newblock


\bibitem[\protect\citeauthoryear{Ipeirotis}{Ipeirotis}{2010}]%
        {ipeirotis2010top}
\bibfield{author}{\bibinfo{person}{Panos Ipeirotis}.}
  \bibinfo{year}{2010}\natexlab{}.
\newblock \showarticletitle{Be a top mechanical turk worker: You need \$5 and 5
  minutes}.
\newblock \bibinfo{journal}{\emph{Blog: Behind Enemy Lines}}
  (\bibinfo{year}{2010}).
\newblock


\bibitem[\protect\citeauthoryear{Jorgensen and Grimstad}{Jorgensen and
  Grimstad}{2005}]%
        {jorgensen2005over}
\bibfield{author}{\bibinfo{person}{Magne Jorgensen} {and}
  \bibinfo{person}{Stein Grimstad}.} \bibinfo{year}{2005}\natexlab{}.
\newblock \showarticletitle{Over-optimism in software development projects:"
  the winner's curse"}. In \bibinfo{booktitle}{\emph{15th International
  Conference on Electronics, Communications and Computers (CONIELECOMP'05)}}.
  IEEE, \bibinfo{pages}{280--285}.
\newblock


\bibitem[\protect\citeauthoryear{Karim, Yang, Messinger, and Ruhe}{Karim
  et~al\mbox{.}}{2018}]%
        {karim2018learn}
\bibfield{author}{\bibinfo{person}{Muhammad~Rezaul Karim}, \bibinfo{person}{Ye
  Yang}, \bibinfo{person}{David Messinger}, {and} \bibinfo{person}{Guenther
  Ruhe}.} \bibinfo{year}{2018}\natexlab{}.
\newblock \showarticletitle{Learn or earn?-intelligent task recommendation for
  competitive crowdsourced software development}.
\newblock  (\bibinfo{year}{2018}).
\newblock


\bibitem[\protect\citeauthoryear{Kaufmann, Schulze, and Veit}{Kaufmann
  et~al\mbox{.}}{2011}]%
        {kaufmann2011more}
\bibfield{author}{\bibinfo{person}{Nicolas Kaufmann}, \bibinfo{person}{Thimo
  Schulze}, {and} \bibinfo{person}{Daniel Veit}.}
  \bibinfo{year}{2011}\natexlab{}.
\newblock \showarticletitle{More than fun and money. Worker Motivation in
  Crowdsourcing-A Study on Mechanical Turk.}. In
  \bibinfo{booktitle}{\emph{Amcis}}, Vol.~\bibinfo{volume}{11}. Detroit,
  Michigan, USA, \bibinfo{pages}{1--11}.
\newblock


\bibitem[\protect\citeauthoryear{Kittur, Nickerson, Bernstein, Gerber, Shaw,
  Zimmerman, Lease, and Horton}{Kittur et~al\mbox{.}}{2013}]%
        {kittur2013future}
\bibfield{author}{\bibinfo{person}{Aniket Kittur}, \bibinfo{person}{Jeffrey~V
  Nickerson}, \bibinfo{person}{Michael Bernstein}, \bibinfo{person}{Elizabeth
  Gerber}, \bibinfo{person}{Aaron Shaw}, \bibinfo{person}{John Zimmerman},
  \bibinfo{person}{Matt Lease}, {and} \bibinfo{person}{John Horton}.}
  \bibinfo{year}{2013}\natexlab{}.
\newblock \showarticletitle{The future of crowd work}. In
  \bibinfo{booktitle}{\emph{Proceedings of the 2013 conference on Computer
  supported cooperative work}}. \bibinfo{pages}{1301--1318}.
\newblock


\bibitem[\protect\citeauthoryear{Lakhani, Garvin, and Lonstein}{Lakhani
  et~al\mbox{.}}{2010}]%
        {lakhani2010topcoder}
\bibfield{author}{\bibinfo{person}{Karim~R Lakhani}, \bibinfo{person}{David~A
  Garvin}, {and} \bibinfo{person}{Eric Lonstein}.}
  \bibinfo{year}{2010}\natexlab{}.
\newblock \showarticletitle{Topcoder (a): Developing software through
  crowdsourcing}.
\newblock \bibinfo{journal}{\emph{Harvard Business School General Management
  Unit Case}} \bibinfo{number}{610-032} (\bibinfo{year}{2010}).
\newblock


\bibitem[\protect\citeauthoryear{Leimeister, Huber, Bretschneider, and
  Krcmar}{Leimeister et~al\mbox{.}}{2009}]%
        {leimeister2009leveraging}
\bibfield{author}{\bibinfo{person}{Jan~Marco Leimeister},
  \bibinfo{person}{Michael Huber}, \bibinfo{person}{Ulrich Bretschneider},
  {and} \bibinfo{person}{Helmut Krcmar}.} \bibinfo{year}{2009}\natexlab{}.
\newblock \showarticletitle{Leveraging crowdsourcing: activation-supporting
  components for IT-based ideas competition}.
\newblock \bibinfo{journal}{\emph{Journal of management information systems}}
  \bibinfo{volume}{26}, \bibinfo{number}{1} (\bibinfo{year}{2009}),
  \bibinfo{pages}{197--224}.
\newblock


\bibitem[\protect\citeauthoryear{Mao, Procaccia, and Chen}{Mao
  et~al\mbox{.}}{2013}]%
        {mao2013better}
\bibfield{author}{\bibinfo{person}{Andrew Mao}, \bibinfo{person}{Ariel
  Procaccia}, {and} \bibinfo{person}{Yiling Chen}.}
  \bibinfo{year}{2013}\natexlab{}.
\newblock \showarticletitle{Better human computation through principled
  voting}. In \bibinfo{booktitle}{\emph{Proceedings of the AAAI Conference on
  Artificial Intelligence}}, Vol.~\bibinfo{volume}{27}.
\newblock


\bibitem[\protect\citeauthoryear{Mao, Yang, Wang, Jia, and Harman}{Mao
  et~al\mbox{.}}{2015}]%
        {mao2015developer}
\bibfield{author}{\bibinfo{person}{Ke Mao}, \bibinfo{person}{Ye Yang},
  \bibinfo{person}{Qing Wang}, \bibinfo{person}{Yue Jia}, {and}
  \bibinfo{person}{Mark Harman}.} \bibinfo{year}{2015}\natexlab{}.
\newblock \showarticletitle{Developer recommendation for crowdsourced software
  development tasks}. In \bibinfo{booktitle}{\emph{2015 IEEE Symposium on
  Service-Oriented System Engineering}}. IEEE, \bibinfo{pages}{347--356}.
\newblock


\bibitem[\protect\citeauthoryear{Mejorado, Saremi, Yang, and
  Ramirez-Marquez}{Mejorado et~al\mbox{.}}{2020}]%
        {mejorado2020study}
\bibfield{author}{\bibinfo{person}{Denisse~Martinez Mejorado},
  \bibinfo{person}{Razieh Saremi}, \bibinfo{person}{Ye Yang}, {and}
  \bibinfo{person}{Jose~E Ramirez-Marquez}.} \bibinfo{year}{2020}\natexlab{}.
\newblock \showarticletitle{Study on Patterns and Effect of Task Diversity in
  Software Crowdsourcing}. In \bibinfo{booktitle}{\emph{Proceedings of the 14th
  ACM/IEEE International Symposium on Empirical Software Engineering and
  Measurement (ESEM)}}. \bibinfo{pages}{1--10}.
\newblock


\bibitem[\protect\citeauthoryear{Richter, Raban, and Rafaeli}{Richter
  et~al\mbox{.}}{2015}]%
        {richter2015studying}
\bibfield{author}{\bibinfo{person}{Ganit Richter}, \bibinfo{person}{Daphne~R
  Raban}, {and} \bibinfo{person}{Sheizaf Rafaeli}.}
  \bibinfo{year}{2015}\natexlab{}.
\newblock \showarticletitle{Studying gamification: the effect of rewards and
  incentives on motivation}.
\newblock In \bibinfo{booktitle}{\emph{Gamification in education and
  business}}. \bibinfo{publisher}{Springer}, \bibinfo{pages}{21--46}.
\newblock


\bibitem[\protect\citeauthoryear{Saremi, Yang, Vesonder, Ruhe, and
  Zhang}{Saremi et~al\mbox{.}}{2021}]%
        {saremi2021crowdsim}
\bibfield{author}{\bibinfo{person}{Razieh Saremi}, \bibinfo{person}{Ye Yang},
  \bibinfo{person}{Gregg Vesonder}, \bibinfo{person}{Guenther Ruhe}, {and}
  \bibinfo{person}{He Zhang}.} \bibinfo{year}{2021}\natexlab{}.
\newblock \showarticletitle{CrowdSim: A Hybrid Simulation Model for Failure
  Prediction in Crowdsourced Software Development}.
\newblock \bibinfo{journal}{\emph{arXiv preprint arXiv:2103.09856}}
  (\bibinfo{year}{2021}).
\newblock


\bibitem[\protect\citeauthoryear{Saremi and Yang}{Saremi and Yang}{2015}]%
        {saremi2015dynamic}
\bibfield{author}{\bibinfo{person}{Razieh~Lotfalian Saremi} {and}
  \bibinfo{person}{Ye Yang}.} \bibinfo{year}{2015}\natexlab{}.
\newblock \showarticletitle{Dynamic simulation of software workers and task
  completion}. In \bibinfo{booktitle}{\emph{2015 IEEE/ACM 2nd International
  Workshop on CrowdSourcing in Software Engineering}}. IEEE,
  \bibinfo{pages}{17--23}.
\newblock


\bibitem[\protect\citeauthoryear{Saremi, Yang, Ruhe, and Messinger}{Saremi
  et~al\mbox{.}}{2017}]%
        {saremi2017leveraging}
\bibfield{author}{\bibinfo{person}{Razieh~L Saremi}, \bibinfo{person}{Ye Yang},
  \bibinfo{person}{Guenther Ruhe}, {and} \bibinfo{person}{David Messinger}.}
  \bibinfo{year}{2017}\natexlab{}.
\newblock \showarticletitle{Leveraging crowdsourcing for team elasticity: an
  empirical evaluation at TopCoder}. In \bibinfo{booktitle}{\emph{2017 IEEE/ACM
  39th International Conference on Software Engineering: Software Engineering
  in Practice Track (ICSE-SEIP)}}. IEEE, \bibinfo{pages}{103--112}.
\newblock


\bibitem[\protect\citeauthoryear{Si and Marsella}{Si and Marsella}{2014}]%
        {si2014encoding}
\bibfield{author}{\bibinfo{person}{Mei Si} {and} \bibinfo{person}{Stacy~C
  Marsella}.} \bibinfo{year}{2014}\natexlab{}.
\newblock \showarticletitle{Encoding theory of mind in character design for
  pedagogical interactive narrative}.
\newblock \bibinfo{journal}{\emph{Advances in Human-Computer Interaction}}
  \bibinfo{volume}{2014} (\bibinfo{year}{2014}).
\newblock


\bibitem[\protect\citeauthoryear{Soliman and Tuunainen}{Soliman and
  Tuunainen}{2015}]%
        {soliman2015understanding}
\bibfield{author}{\bibinfo{person}{Wael Soliman} {and}
  \bibinfo{person}{Virpi~Kristiina Tuunainen}.}
  \bibinfo{year}{2015}\natexlab{}.
\newblock \showarticletitle{Understanding continued use of crowdsourcing
  systems: An interpretive study}.
\newblock \bibinfo{journal}{\emph{Journal of theoretical and applied electronic
  commerce research}} \bibinfo{volume}{10}, \bibinfo{number}{1}
  (\bibinfo{year}{2015}), \bibinfo{pages}{1--18}.
\newblock


\bibitem[\protect\citeauthoryear{Stewart, Huerta, and Sader}{Stewart
  et~al\mbox{.}}{2009}]%
        {stewart2009designing}
\bibfield{author}{\bibinfo{person}{Osamuyimen Stewart}, \bibinfo{person}{Juan~M
  Huerta}, {and} \bibinfo{person}{Melissa Sader}.}
  \bibinfo{year}{2009}\natexlab{}.
\newblock \showarticletitle{Designing crowdsourcing community for the
  enterprise}. In \bibinfo{booktitle}{\emph{Proceedings of the ACM SIGKDD
  Workshop on Human Computation}}. \bibinfo{pages}{50--53}.
\newblock


\bibitem[\protect\citeauthoryear{Terveen and Hill}{Terveen and Hill}{2001}]%
        {terveen2001beyond}
\bibfield{author}{\bibinfo{person}{Loren Terveen} {and} \bibinfo{person}{Will
  Hill}.} \bibinfo{year}{2001}\natexlab{}.
\newblock \showarticletitle{Beyond recommender systems: Helping people help
  each other}.
\newblock \bibinfo{journal}{\emph{HCI in the New Millennium}}
  \bibinfo{volume}{1}, \bibinfo{number}{2001} (\bibinfo{year}{2001}),
  \bibinfo{pages}{487--509}.
\newblock


\bibitem[\protect\citeauthoryear{Whitehill, Wu, Bergsma, Movellan, and
  Ruvolo}{Whitehill et~al\mbox{.}}{2009}]%
        {whitehill2009whose}
\bibfield{author}{\bibinfo{person}{Jacob Whitehill}, \bibinfo{person}{Ting-fan
  Wu}, \bibinfo{person}{Jacob Bergsma}, \bibinfo{person}{Javier Movellan},
  {and} \bibinfo{person}{Paul Ruvolo}.} \bibinfo{year}{2009}\natexlab{}.
\newblock \showarticletitle{Whose vote should count more: Optimal integration
  of labels from labelers of unknown expertise}.
\newblock \bibinfo{journal}{\emph{Advances in neural information processing
  systems}}  \bibinfo{volume}{22} (\bibinfo{year}{2009}),
  \bibinfo{pages}{2035--2043}.
\newblock


\bibitem[\protect\citeauthoryear{Yang, Karim, Saremi, and Ruhe}{Yang
  et~al\mbox{.}}{2016}]%
        {yang2016should}
\bibfield{author}{\bibinfo{person}{Ye Yang}, \bibinfo{person}{Muhammad~Rezaul
  Karim}, \bibinfo{person}{Razieh Saremi}, {and} \bibinfo{person}{Guenther
  Ruhe}.} \bibinfo{year}{2016}\natexlab{}.
\newblock \showarticletitle{Who should take this task? Dynamic decision support
  for crowd workers}. In \bibinfo{booktitle}{\emph{Proceedings of the 10th
  ACM/IEEE International Symposium on Empirical Software Engineering and
  Measurement}}. \bibinfo{pages}{1--10}.
\newblock


\bibitem[\protect\citeauthoryear{Yang and Saremi}{Yang and Saremi}{2015}]%
        {yang2015award}
\bibfield{author}{\bibinfo{person}{Ye Yang} {and} \bibinfo{person}{Razieh
  Saremi}.} \bibinfo{year}{2015}\natexlab{}.
\newblock \showarticletitle{Award vs. worker behaviors in competitive
  crowdsourcing tasks}. In \bibinfo{booktitle}{\emph{2015 ACM/IEEE
  International Symposium on Empirical Software Engineering and Measurement
  (ESEM)}}. IEEE, \bibinfo{pages}{1--10}.
\newblock


\bibitem[\protect\citeauthoryear{Ye, Wang, and Liu}{Ye et~al\mbox{.}}{2015}]%
        {ye2015crowd}
\bibfield{author}{\bibinfo{person}{Bin Ye}, \bibinfo{person}{Yan Wang}, {and}
  \bibinfo{person}{Ling Liu}.} \bibinfo{year}{2015}\natexlab{}.
\newblock \showarticletitle{Crowd trust: A context-aware trust model for worker
  selection in crowdsourcing environments}. In \bibinfo{booktitle}{\emph{2015
  IEEE International Conference on Web Services}}. IEEE,
  \bibinfo{pages}{121--128}.
\newblock


\bibitem[\protect\citeauthoryear{Yin, Chen, and Sun}{Yin et~al\mbox{.}}{2014}]%
        {yin2014monetary}
\bibfield{author}{\bibinfo{person}{Ming Yin}, \bibinfo{person}{Yiling Chen},
  {and} \bibinfo{person}{Yu-An Sun}.} \bibinfo{year}{2014}\natexlab{}.
\newblock \showarticletitle{Monetary interventions in crowdsourcing task
  switching}. In \bibinfo{booktitle}{\emph{Second AAAI Conference on Human
  Computation and Crowdsourcing}}.
\newblock


\bibitem[\protect\citeauthoryear{Zhang, Wu, and Wu}{Zhang
  et~al\mbox{.}}{2015}]%
        {zhang2015analyzing}
\bibfield{author}{\bibinfo{person}{Hui Zhang}, \bibinfo{person}{Yuchuan Wu},
  {and} \bibinfo{person}{Wenjun Wu}.} \bibinfo{year}{2015}\natexlab{}.
\newblock \showarticletitle{Analyzing developer behavior and community
  structure in software crowdsourcing}.
\newblock In \bibinfo{booktitle}{\emph{Information science and applications}}.
  \bibinfo{publisher}{Springer}, \bibinfo{pages}{981--988}.
\newblock


\end{thebibliography}

\end{document}